\documentclass[12pt]{article}
\usepackage{mathrsfs,bm,color}
\usepackage{amsmath}
\usepackage{amssymb}

\newtheorem{theorem}{Theorem}[section]
\newtheorem{df}[theorem]{\bf Definition}
\newtheorem{thm}[theorem]{\bf Theorem}

\newtheorem{lem}[theorem]{\bf Lemma}

\newtheorem{prop}[theorem]{\bf Proposition}
\newtheorem{assumption}[theorem]{\bf Assumption}
\numberwithin{equation}{section}
\newcommand{\proof}{{\noindent \it Proof:\ }}

\newcommand{\ix}
{\int_{\BR}\!\!\! {\rm d}x\Ebb_{P\times\nu}^x}
\newcommand{\iix}
{\int_{\BR}\!\!\! {\rm d}x\Ebb_{P}^x}
\newcommand{\la}{\lambda}

\newcommand{\A}{\ms A}
\renewcommand{\AA}{{\ms A_{E}}}
\renewcommand{\j}{{\rm j}}
\renewcommand{\Re}{{\rm Re}}

\newcommand{\qed}{\hfill\hbox{\rule{6pt}{6pt}}}

\newcommand{\eps}{\epsilon}
\newcommand{\ov}[1]{\overline{#1}}

\newcommand{\lk}{\left(\!}
\newcommand{\rk}{\!\right)}
\newcommand{\lkk}{\left\{\!}
\newcommand{\rkk}{\!\right\}}

\newcommand{\BR}{\RR^d}
\newcommand{\RR}{\mathbb R}

\newcommand{\rel}{V_{\rm rel}}
\newcommand{\conf}{V_{\rm conf}}

\renewcommand{\d}{\displaystyle}
\newcommand{\add}{a^{\dagger}}
\newcommand{\ass}{a^\sharp}

\newcommand{\LR}{{L^2(\BR)}}
\newcommand{\hp}{{\rm H}_{\rm p}}
\newcommand{\Ebb}{\mathbb E}

\newcommand{\fff}{{\mathscr{F}}}
\newcommand{\hhh}{{\mathscr{H}}}
\newcommand{\ffff}{{\mathscr{F}_{\rm fin}}}
\newcommand{\hhhh}{{\mathscr{H}_{\rm fin}}}

\newcommand{\hf}{{\rm H}_{\rm f}}
\newcommand{\pf}{{\rm P}_{\rm f}}

\newcommand{\tot}{{{\rm P}_{\rm tot}}}

\newcommand{\half}{\frac{1}{2}}

\newcommand{\han}{{1/2}}

\newcommand{\non}{\nonumber}
\newcommand{\bi}{\begin{description}}
\newcommand{\ei}{\end{description} }

\def\bbbone{{\mathchoice {\rm 1\mskip-4mu l} {\rm 1\mskip-4mu l}
{\rm 1\mskip-4.5mu l} {\rm 1\mskip-5mu l}}}
\def\one{\bbbone}

\newcommand{\vp}{\hat\varphi}

\providecommand{\eq}[1]{\begin{equation}\label{#1}}
\providecommand{\en}{\end{equation}}

\providecommand{\kak}[1]{(\ref{#1})}
\providecommand{\HHH}{\ms H}
\providecommand{\ms}[1]{\mathscr{#1}}
\providecommand{\tmm}{T_m}
\providecommand{\T}{T_M}

\title{\sc Self-adjointness of the semi-relativistic Pauli-Fierz Hamiltonian}
\author{\small
Takeru Hidaka
\thanks{
Faculty of Mathematics,
Kyushu University, Motooka 744, Nishiku, Fukuoka 819-0385, Japan.
hidaka@math.kyushu-u.ac.jp.}
 and  Fumio Hiroshima
\footnote{
Faculty of Mathematics,
Kyushu University, Motooka 744, Nishiku,  Fukuoka,  819-0385, Japan. 
hiroshima@math.kyushu-u.ac.jp.
This work is partially supported by 
 Grant-in-Aid for Science Research (B) 23340032. 
}
}
\date{\today}
\begin{document}
\maketitle
\begin{abstract}
The spinless semi-relativistic Pauli-Fierz Hamiltonian 
$$H=\sqrt{(p\otimes \one -A)^2+M^2}+V\otimes \one +\one \otimes \hf, $$
in quantum electrodynamics 
is considered. Here $p$ denotes a momentum operator, $A$ a quantized radiation field, $M\geq 0$, $\hf$ the free hamiltonian of a Boson Fock space and $V$ 
an external potential.  
The self-adjointness and essential self-adjointness  of $H$ 
are shown.  It is emphasized that it includes the case of $M=0$.  
Furthermore,   the self-adjointness and the essential self-adjointness 
of the semi-relativistic 
Pauli-Fierz model with a fixed total momentum $P\in \BR$: 
$$H(P)=\sqrt{(P-\pf-A(0))^2+M^2}+\hf, \quad M\geq 0,$$
is also proven for arbitrary $P$. 
\end{abstract}

\section{Introduction}
\subsection{Fundamental facts}
In this paper we are concerned with the self-adjointness of the so-called semi-relativistic Pauli-Fierz (SRPF) Hamiltonian $H$ 
in quantum electrodynamics. 
Essential self-adjointness of $H$ 
is shown in \cite[Theorem 4.5]{hir14} by a path measure approach under some conditions.  
We furthermore show its self-adjointness 
under weaker conditions in this paper. Our result is independent of coupling constants. 
In this sense the result is non-perturbative. 

Let $\ms K$ be a Hilbert space over $\Bbb C$ and 
$h$ be  a symmetric  operator with the domain  $D_0$. 
In general $h$ has the infinite number of self-adjoint extensions. 
Let $h_0$ be one self-adjoint extension, which defines the Schr\"odinger equation 
\eq{sch}
i\frac{\partial}{\partial t} \Phi_t=h_0 \Phi_t
\en
 with the initial condition $\Phi_0=
 \Phi\in \ms K$. 
 Then the self-adjointness of $h_0$ ensures the uniqueness of the solution to \kak{sch} and it is given by $\Phi_t=e^{-ith_0}\Phi$. 
The time-evolution of a physical system governed by the Schr\"odinger equation \kak{sch}  is different according to which self-adjoint extension is chosen.
Hence it is important to find a core  
of $h$ or a domain on which $h$ is self-adjoint 
in order to determine the unique time-evolution of the physical system.

A semi-relativistic Schr\"odinger operator 
with nonnegative rest mass $M\geq 0$ is defined as a self-adjoint operator in $\LR$, which is  
given by
\eq{relativistic}
\hp=\sqrt{p^2+M^2}+V.
\en
Here $p=(-i\partial_{x_1},\cdots,-i\partial_{x_d})$ denotes the momentum operator 
and $V:\BR\to \RR$ is 
an external potential.
The SRPF model is defined by $\hp$ coupled to a quantized radiation field $A$. 
Let $\fff=
\oplus_{n=0}^\infty 
\fff_n(W)=
\oplus_{n=0}^\infty 
\otimes_s^n W$
be the Boson Fock space over
Hilbert space 
$W=\oplus^{d-1}\LR$, $d\geq 3$.  
Although the physically reasonable choice of the spatial dimension 
is $d=3$, we generalize it.
Let $\omega:\BR\to \RR$ be a dispersion relation. 
We introduce assumptions on the dispersion relation.
\begin{assumption}
\label{h0}
$\omega(k)\geq0$ a.e. $k\in\BR$.
\end{assumption}
Physically reasonable choice of dispersion relation is $\omega(k)=|k|$ or $\omega(k)=\sqrt{|k|^2+\nu^2}$ with some $\nu>0$.  
In \cite{hh13} the dispersion relation 
such that 
$\omega\in C^1 (\BR; \RR)$, 
$\nabla\omega\in L^\infty(\BR)$, 
 $\d \inf_{k\in\BR } \omega(k)\geq m$ 
with some $m>0$ and 
$\d \lim_{|k|\to\infty}\omega(k)=\infty$ 
is treated. 
The free field Hamiltonian $H_{\rm f}$ of the
Boson Fock space is given by the second quantization of 
the multiplication operator 
by $\omega$ on $W$, i.e., 
$\hf =d\Gamma(\omega)$. 
The SRPF Hamiltonian is defined by the minimal coupling of a quantized radiation field $A$ 
to 
\eq{Hzero}
H_0=\hp\otimes\one+\one\otimes\hf.
\en
$H_0$ is self-adjoint on 
$D(\hp\otimes\one)\cap D(\one\otimes \hf)$. 
The creation operator and the annihilation operator in $\fff$ are denoted by 
$\add(f)$ and $a(f)$, $f\in W$, respectively. 
They 
are 
 linear in  $f$ and 
satisfy 
canonical commutation relations: 
$[a(f),\add(g)]=(\bar f , g)_W$ and 
$ [a(f),a(g)]=0=[\add(f),\add(g)]$.
Here and in what follows 
the scalar product $(f,g)_{\mathscr K}$ 
on a Hilbert space $\mathscr K$ is linear in 
$g$ and anti-linear in $f$. 
We formally write as $\d a^{\# r}(f)=\int a^{\# r}(k) f(k) {\rm d}k$ for $\ass(F)$ with $F=\oplus_{s=1}^{d-1}\delta_{sr}f$ and 
$$\d \hf=\sum_{r=1}^{d-1}\int \omega(k) a^{\dagger r}(k) a^r(k) {\rm d}k.$$
Let $e^r(k)=(e^r_1(k),...,e^r_d(k))$ be 
$d$-dimensional polarization vectors, i.e., 
$e^r(k)\cdot e^s(k)=\delta _{rs}$ and $k\cdot e^r(k)=0$ for $k\in\BR\setminus\{0\}$ and  $r=1,...,d-1$. 
For  each $x\in\BR$ a quantized radiation field 
$A(x)=(A_1(x),...,A_d(x))$ 
is 
defined  by 
\begin{eqnarray}
A_\mu(x)=\frac{1}{\sqrt 2}
\sum_{r=1}^{d-1}\int e_\mu^r(k)
\left(\frac{\hat\varphi(k)  e^{-ik\cdot x}}{\sqrt{\omega(k)}} a^{\dagger r}(k) + \frac{\hat\varphi(-k)  e^{ik\cdot x}}{\sqrt{\omega(k)}} a^r(k) \right){\rm d}k.
\end{eqnarray}
Here  $\vp$ is  an ultraviolet cutoff function, 
for which we introduce assumptions below.
\begin{assumption}\label{h1}
$\vp/\sqrt\omega, 
\omega\sqrt\omega\vp \in\LR$ 
and $\vp(k)=\ov{\vp(-k)}$.
\end{assumption}
Note that $\sqrt\omega\vp\in\LR$ follows from Assumption \ref{h1}. 
We fix $\vp$ and $\omega$ satisfying 
Assumptions \ref{h0} and  
\ref{h1} throughout this paper. 
Then 
 $\vp(k)=\ov{\vp(-k)}$ implies that  $A_\mu(x)$ is essentially self-adjoint for each $x$. 
 We denote the self-adjoint extension by the same symbol $A_\mu(x)$. 
We identify $\hhh$ with 
$\d \int^\oplus_{\BR} \!\!\!\fff {\rm d}x$, 
and under this identification we 
define  the self-adjoint operator 
$A_\mu$ in $\hhh$ by 
$$A_\mu=
\int_{\BR}^\oplus\!\!\! A_\mu (x) {\rm d}x.$$  
Set $A=(A_1,\cdots,A_d)$. 
Let $N=d\Gamma(\one)$ be the number operator on $\fff$ and 
 $C^\infty(\one\otimes N)=
\cap_{n=1}^\infty 
D(\one\otimes N^n)$. 
Let 
\eq{pa}
\sum_{\mu=1}^d 
(p_\mu\otimes\one-A_\mu)^2=
(p\otimes \one -A)^2.
\en
\begin{lem}
\label{hiroshima}
$D(p^2\otimes \one )\cap 
C^\infty(\one\otimes N)\cap D(\one\otimes\hf)
$ is a core of 
$(p\otimes\one -A)^2$.
\end{lem}
\proof See  Appendix \ref{phiroshima}. 
\qed

The closure of 
$(p\otimes\one -A)^2
\lceil_{D(p^2\otimes \one )
\bigcap C^\infty(\one\otimes N)\bigcap D(\one\otimes\hf)}$ is denoted by 
$(p\otimes\one -A)^2$ for simplicity.
Thus 
$\sqrt{(p\otimes \one -A)^2+M^2}$ is defined through the spectral measure 
of $(p\otimes\one -A)^2$.
 Set 
 \eq{tt}
\T=\sqrt{(p\otimes \one -A)^2+M^2}.
 \en
\begin{prop}{\rm \cite[Lemma 3.12, Theorem 4.5]{hir14}}
\label{hiro10}
Let
 $M>0$. 
Then (1) and (2) follow. 
\bi
\item[(1)]
Let $V=0$. Then $H$ is essentially self-adjoint on 
$\ms D$.
\item[(2)] Suppose that $V$ is relatively 
 bounded (resp. form bounded) 
 with respect to $\sqrt{\!p^2+M^2\!}$  
 with a relative bound $a$. 
Then $V$ is also relatively bounded (resp. form bounded) with respect to 
$\T +\hf$ with a relative bound smaller than $a$. 
\ei
\end{prop}
\subsection{Potential classes and definition of SRPF Hamiltonian}
We introduce two classes, 
$V_{\rm qf}$ and 
$V_{\rm rel}$,  
 of potentials.
\begin{df} 
{\rm 
($V_{\rm qf}$) 
$V=V_+-V_-\in V_{\rm qf}$ if and only if 
$V_+\in L_{loc}^1(\BR)$ and $V_-$ is relatively form bounded with respect to 
$\sqrt{p^2+M^2}$ with a relative bound strictly smaller than one, i.e., 
$D((p^2+M^2)^{1/4})\subset D(V_-^\han)$ and 
there exist $0\leq a<1$ and $b\geq 0$ such that 
$$\|V_-^\han f\|\leq 
a\|({p^2+M^2})^{1/4} f\|+b\|f\|$$
for all $f\in D(({p^2+M^2})^{1/4})$. \\
($V_{\rm rel}$)
$V\in V_{\rm rel}$ if and only if 
$V$ is relatively bounded with respect to 
$\sqrt{p^2+M^2}$ with a relative bound strictly 
smaller than one, i.e.,  
$D(\sqrt{p^2+M^2})\subset D(V)$ and 
there exist $0\leq a<1$ and $b\geq 0$ such that 
$$\|V f\|\leq 
a\|\sqrt{p^2+M^2} f\|+b\|f\|$$
for all $f\in D(\sqrt{p^2+M^2})$. 
}\end{df}
It can be shown that 
$V_{\rm rel}\subset V_{\rm qf}$.
By Proposition \ref{hiro10}
we can define the SRPF Hamiltonian as a self-adjoint operator through  
quadratic form sums.
 Let $V\in V_{\rm qf}$. 
We define the quadratic form by 
\eq{qf1}
q:(F, G)\mapsto (\T^\han  F, \T^\han  G)+(\hf^\han F, \hf^\han G)+(V^\han_+F, V^\han_+ G)-(V^\han_- F, V^\han_- G)
\en 
with the form domain 
\eq{qf2}
Q(q)=
D(\T^\han )\cap D(\hf^\han)\cap D(V_+^\han).
\en
By Proposition \ref{hiro10}, 
we note that 
$Q(q)
=D(\T^\han )\cap D(\hf^\han)\cap D(V_+^\han)\cap D(V_-^\han)$.
It can be checked that 
$Q(q)$ is densely defined semi-bounded closed form. 
Then there exists the unique  self-adjoint operator $H$ associated with 
the quadratic form $q$, i.e., $D(|H|^\han)=Q(q)$ and 
$q(F, G)=\int_{\sigma(H)} \la d(E_\la F, G)$. Here $E_\la$ denotes the spectral measure associated with $H$.  
We write $H$ as 
\begin{align}
\label{srpf}
H =\T
\ \dot +\ 
V_+\otimes \one\ \dot-\ V_-\otimes\one\ \dot  +\ \one\otimes \hf. 
\end{align}

\begin{df}
Let $V\in V_{\rm qf}$. Then the 
SRPF  Hamiltonian is defined by 
\kak{srpf}. 
\end{df}
We do not write tensor notation $\otimes$ for 
notational convenience in what follows. 
Thus 
$H$ can be simply written as 
$
H=\T \ \dot + \ \hf\  \dot +\ V_+\ \dot -\ V_-$.
\subsection{Essential self-adjointness of $H$}
Let 
\eq{D}
\ms D=D(|p|)\cap D(V)\cap D(\hf).
\en
When $V\in V_{\rm rel}$, 
$D(V)\subset D(|p|)\cap D(\hf)$ and 
it follows that  
$\ms D=D(|p|)\cap D(\hf)$. 
We introduce a subclass 
$V_{\rm conf}\subset V_{\rm qf}$, which include confining potentials. 
\begin{df}{\rm ($\conf$)}
\label{h2}
$V=V_+-V_- \in V_{\rm conf}$ if and only if $V_-=0$ and $V_+$  is  twice differentiable, 
 and $\partial_\mu V_+, 
 \partial_\mu^2 V_+ \in L^\infty(\BR)$
  for 
   $\mu=1,...,d$, 
and  $D(V)\subset D(|x|)$. 
\end{df}
When $V\in V_{\rm conf}$, 
$V\in L_{loc}^2(\BR)$ and nonnegative. Then 
$p^2+V$ is essentially self-adjoint on $C_{\rm c}^\infty(\BR)$  by Kato's inequality. 
It is established in \cite[Theorem 4.5]{hir14} 
that $H$ with $M>0$ 
is essentially self-adjoint on $\ms D$ for $V\in V_{\rm rel}$.  We extend this to 
$V\in V_{\rm rel}\cup \conf$. 
\begin{prop}
\label{hiro11}
Let $V\in V_{\rm rel}\cup \conf$
and $M>0$. 
Then 
$H$ is essentially self-adjoint on  $\ms D$.
\end{prop}
\proof 
When $V\in \rel$, 
the proposition follows from (2) of Proposition \ref{hiro10} and  the Kato-Rellich theorem.
The proof of the proposition for $V\in \conf$  is a minor modification of \cite[Theorem 4.5]{hir14}. 
Then we  give it  in Appendix \ref{phiro11}.  
\qed

\subsection{Main results}
The self-adjointness of the Pauli-Fierz Hamiltonian:  
$$\half (p\otimes \one -\alpha A)^2+V\otimes \one+\one\otimes\hf$$
is proven in \cite{hh08,hir00,hir02} for arbitrary values of coupling constant $\alpha\in\RR$ under some condition on $\vp$ and $V$.  
On the other hand 
as far as we know 
there a few work on 
the self-adjointness of the SRPF Hamiltonian.
In \cite{ms09} the self-adjointness of the SRPF Hamiltonian
with spin $\han$ and without $V$: 
$$\gamma \sqrt{(\sigma\cdot (p\otimes \one-\alpha A))^2+M^2}+\one\otimes\hf$$
is  shown  for $d=3$ but for 
sufficiently small coupling constant $\alpha$,
where $\sigma=(\sigma_1,\sigma_2, \sigma_3)$ denotes  $2\times 2$ Pauli matrices and $0<\gamma\leq 1$ an artificial parameter. 
In \cite{ms09} the self-adjointness is proven by a perturbation theory, i.e., 
operator  $|D|-|D_0|$ is estimated for sufficiently small $\alpha$,  where 
 $|D|=\sqrt{(\sigma\cdot (p\otimes \one-\alpha A))^2+M^2}$ and 
 $|D_0|=\sqrt{(\sigma\cdot p)^2+M^2}$, and 
 the self-adjointness of $|D|+\hf$ can be reduce to show that of $|D_0|+\hf$ for sufficiently small $\alpha$.  
This is unfortunately not applicable for arbitrary values of $\alpha$. 
By functional integration  however it is proven  in \cite{hir14} that $H$ is essentially self-adjoint on $\ms D$ for $M>0$, which is due to 
show  that $e^{-tH}\ms D\subset \ms D$.    

Then  the main purpose of this paper is to show the self-adjointness of $H$ on $\ms D$ 
for arbitrary values of coupling constants 
(in this paper $\alpha$ is absorbed in the prefactor of $\vp$), and 
not only for $V\in \rel$ but also  for $V\in \conf$.  
This can be achieved  by proving the nontrivial bound 
\kak{atari2} mentioned below, which bound implies the closedness of $H\lceil_{\ms D}$. 
 In order to prove \kak{atari2} for $0\leq M$ we have to estimate 
the commutator like  $[\sqrt{(p-A)^2+M^2}, \cdot]$. See the proof of Lemma \ref{relative bound}. 
In particular the proof for the case of $M=0$ is not technically straightforward, and then we used a functional integral method.

We define the dense subset 
$\ms H_{\rm fin}$. 
Let 
\eq{yasiki2}
\ffff=L.H.\{\Omega, a^\dagger (h_1)\cdots a^\dagger(h_n) \Omega | h_j \in C_{\rm c}^\infty (\BR), j=1,\cdots,n, n\geq 1\}
\en
and  
\eq{yasiki}
{\ms H}_{\rm fin}=C_{\rm c}^\infty(\BR)
\widehat \otimes \ffff,
\en
where $\widehat \otimes$ denotes the algebraic tensor product. 
The main theorem in this paper is to extend 
Proposition \ref{hiro11} as follows. 
\begin{thm}
\label{main}
Let $V\in \rel\cup \conf$
and $M\geq 0$. 
Then 
$H$  is self-adjoint on 
$\ms D$,   
and  essentially self-adjoint on 
$\HHH_{\rm fin}$.
\end{thm}
Note that Theorem \ref{main} includes 
the case of $M=0$.

\subsection{Literatures and organization}
We refer to literatures where the SRPF model is studied. 
In \cite{gll01} the existence of the ground state of the SRPF model is suggested and the present work is inspired from this.    
Then the ground state of the SRPF model is studied in e.g., \cite{gs12, hh13, hir14, km13a, km13b, km14,kms11a, kms11b,ms10,ms09},
in particular the  case of $M=0$ is investigated in \cite{hh13}. 
Moreover in \cite{fgs01} the asymptotic analysis of the SRPF model is also studied.

This paper is organized as follows. 

In Section 2 we show that $H$ is self-adjoint on $D(|p|)\cap D(V)\cap D(\hf)$ and essentially self-adjoint on $\hhhh$ which is defined in \kak{yasiki}.

In Section 3 we discuss the translation invariant SRPF Hamiltonian which is defined by $H$ with $V=0$. 
Then $H\cong \int^\oplus _{\BR} H(P) dP$ is obtained and $H(P)$ is called the SRPF Hamiltonian with total momentum $P\in\BR$. 
The self-adjointness of $H(P)$ on $D(|\pf|)\cap D(\hf)$, and essential self-adjointness on $\ffff$ defined in \kak{yasiki2}.
   
\section{Self-adjointness}

In order to prove Theorem \ref{main} we need several lemmas.

\begin{lem}
Let $M\geq 0$. It follows that 
$D(|p|) \cap D(\hf^{1/2})  \subset D\left( 
\T \right)$, 
and for all $\Psi\in D(|p|)\cap D(\hf^{1/2})$,
\eq{atari}
\left\| \T  \Psi \right\| \leq 
C(\||p|\Psi\|+\|\hf^\han\Psi\|+\|\Psi\|)
\en
with some constant $C>0$.
In particular  
\eq{atari2}
\left\| H  \Psi \right\| \leq 
C(\||p|\Psi\|+\|\hf \Psi\|+\|V\Psi\|+\|\Psi\|)
\en
follows for $\Psi\in \ms D$ with some constant $C>0$.
\end{lem}
\proof
It follows that 
$
\|\T \Psi\|^2=
\sum_{\mu=1}^d\|(p_\mu-A_\mu)\Psi\|^2+M^2\|\Psi\|^2$
for 
$\Psi\in 
\hhhh$. 
Then 
\kak{atari}
follows from the well-known bound 
$$\|A_\mu \Psi\|\leq C\|(\hf+\one)^\han\Psi\|$$ with some constant $C>0$ 
for 
$\Psi\in \hhhh$. 
Furthermore since 
$|p|+\hf^\han$ is essentially self-adjoint on 
$\hhhh$,
the lemma  follows from a limiting argument. 
\qed

Let
$
\ms H_0 =\lkk 
\{\Psi^{(n)}\}_{n=0}^\infty 
\in\ms H | \Psi^{(n)}=0
 \mbox{ for all } n\geq n_0 \mbox{ with some } n_0\geq 1 \rkk$ 
and 
\eq{d1}
\ms D_1=\ms D\cap
\ms H_0.
\en
\begin{lem}
\label{lemma1}
Let $V\in\conf$ and $M>0$.
Then 
$\ms D_1$  is a core of  $H$.  
\end{lem}
\proof
Let $P_n=\one_{[0,n]}(N)$ for $n\in \mathbb N$. 
Take an arbitrary 
$\Psi\in \ms D$.
Hence $P_n\Psi\in \ms D_1$. 
 We see that
$P_n\Psi \to \Psi$ as $n\to\infty$. 
Since
\begin{align}
\|  H(P_n-P_{n'})\Psi \|
 \leq 
 C(
 \|(P_n-P_{n'})|p|\Psi\|+
\| (P_n-P_{n'}) V\Psi\|+
\| (P_n-P_{n'}) \hf\Psi\|) ,
\end{align}
we also see that $\{H P_n\Psi\}_n$ is a Cauchy sequence in $\HHH$. 
By the closedness of $H$, $\Psi\in D(H)$ and 
$HP_n\Psi\to H\Psi$. Thus $\ms D_1$ is a core of $H$. 
\qed

Let
\eq{d2}
\ms D_2=\{\{\Psi^{(n)}\}_{n=0}^\infty \in \ms D_1 | \Psi^{(n)}
(\cdot, {\bf k})\in C_{\rm c}^\infty (\BR)    \text{ a.e. }   {\bf k}\in \RR^{dn},  n\geq 1 \}.
\en
\begin{lem}
Let $V\in\conf$ and $M>0$.
Then $\ms D_2$ is a core of  $H$.
\end{lem}
\proof
Take an arbitrary 
$\Phi\in \ms D_1
$.
Let
$j\in C_{\rm c}^\infty(\BR)$
 and $g\in C_{\rm c}^\infty (\BR;[0,1])$ such that 
 $\int_{\BR} j(x){\rm d}x=1$ and $g(x)=1$ for $ |x|\leq 1$.
For each $\epsilon>0$ we set $j_\epsilon (x)=\epsilon^{-d}j(x/\epsilon)$,
\begin{eqnarray}
\Phi_{\epsilon,L}^{(n)}(x,{\bf k})= 
g(x/L) \int_{\BR} j_\epsilon(x-y) \Phi^{(n)}(y,{\bf k}) dy,
\end{eqnarray}
and $\Phi_{\epsilon,L}=\{\Phi_{\epsilon,L}^{(n)}\}_{n=0}^\infty $.
We see that $\Phi_{\epsilon,L}\to\Phi$,
$p_\mu\Phi_{\epsilon,L}\to p_\mu\Phi$, $V\Phi_{\epsilon,L}\to V\Phi$ and $\hf\Phi_{\epsilon,L}\to\hf\Phi$ 
strongly as 
$\epsilon\downarrow 0$ and $L\to\infty$.
Then by inequality \kak{atari2}
and the closedness of $H$, 
we see that 
$\Phi\in D(H)$ and 
$
\d \lim_{L\to\infty}
\lim_{\eps\downarrow 0} H\Phi_{\epsilon,L} 
= H\Phi$ 
in $\HHH$. Thus the lemma follows. 
\qed

\begin{lem}
Let $V\in\conf$ and $M>0$.
Let $\Phi  \in \ms D_2$. 
Then 
it follows that 
\eq{hidaka1}
\|  p^2\Phi  \| +\|  V\Phi \| + \|  \hf\Phi  \|  \leq C\|  (p^2+V+\hf+\one)\Phi  \| 
\en
with some constant $C>0$.
\end{lem}
\proof
Note that 
$
\|(p^2+V)\Phi \|^2=\|p^2\Phi \|^2+
2\Re(p^2\Phi , V\Phi )+
\|V\Phi \|^2
$. 
Let $V_\mu=\partial_\mu V$. 
Since 
\begin{align*}
2\Re(p^2\Phi , V\Phi )
&\geq
2\sum_\mu \Re(p_\mu \Phi , 
[p_\mu, V] \Phi )\geq 
-2\sum_\mu  \|p_\mu\Phi\| 
\|V_\mu\| _\infty  \|\Phi \|,
\end{align*}
for an arbitrary $\eps>0$, we have 
$
\|(p^2+V)\Phi \|^2\geq 
(1-\eps)\|p^2\Phi \|^2+\|V\Phi \|^2-C_\eps\|\Phi \|^2
$
and 
\begin{align*}
\|(p^2+V+\hf)\Phi \|^2
&\geq 
\|(p^2+V)\Phi \|^2+
\|\hf \Phi \|^2\\
&\geq 
(1-\eps)\|p^2\Phi \|^2+\|V\Phi \|^2-C_\eps\|\Phi \|^2+\|\hf\Phi \|^2.
\end{align*}
Then \kak{hidaka1} follows. 
\qed

\begin{lem}
\label{lemma2}
Let $V\in\rel\cup \conf$ and $M\geq 0$.
Then $\hhhh$ is a core of $H$.
\end{lem}
\proof
Suppose that $M>0$. 
Let $\Phi\in \ms D_2$. 
Let $V\in \conf$. 
Note that $p^2+V+\hf$ is essentially self-adjoint on $\HHH_{\rm fin}$.
We see that there exists a sequence 
$\{\Phi _n\}$, $\Phi _n\in \HHH_{\rm fin}$, 
such that
$\Phi _n\to\Phi $, and $(p^2+V+\hf )\Phi _n\to (p^2+V+\hf)\Phi $ as $n\to \infty$. 
From  \kak{hidaka1} it follows that 
$p^2 \Phi _n\to p^2 \Phi $, 
$V\Phi _n\to V\Phi $ and 
$\hf\Phi _n\to\hf\Phi $ as $n\to\infty$. 
Then we can also see that  
$\{H\Phi _n\}$ is a Cauchy sequence by \kak{atari2}, 
and 
$\lim_n H\Phi _n=
H\Phi $ follows. 
Thus $\HHH_{\rm fin}$ is a core of $H$.
Next we suppose that $V\in \rel$. 
By the argument above it is seen that operator $\T \ \dot + \ \hf$ is essentially self-adjoint on $\hhhh$. 
By Proposition \ref{hiro10}, we also see that 
$\|V\Phi\|\leq a \|(\T \ \dot + \ \hf)\Psi\|+b\|\Psi\|$
with some constant $0\leq a<1$ and $b\geq 0$. The Kato-Rellich theorem yields that $H$ 
is essentially self-adjoint on $\hhhh$.

Suppose that $M=0$.
We emphasis the dependence on $M$ by writing $H_M$ for $H$. 
Since 
$H_0=H_M+(H_0-H_M)$ and 
$
\| (H_0-H_M)\Psi\| \leq M\| \Psi\|$,
$H_0$ is also essentially self-adjoint on 
$\hhhh$ by  the fact that $H_M$ is essentially self-adjoint on $\hhhh$ and by the Kato-Rellich theorem. 
\qed

The key inequality to show 
the self-adjointness of $H$ on $\ms D$ is the following  inequality. 

\begin{lem} \label{relative bound}
Let $V\in\conf$. 
Let $M_0>0$ be fixed and $0\leq M\leq M_0$. 
Then for all $\Psi\in D(H)$,
\eq{hiro2015}
\|  |p| \Psi \| ^2 + \|  V\Psi \| ^2 +\|  \hf \Psi \| ^2  \leq C \|  (H+\one) \Psi \| ^2
\en
with some constant $C$ independent of $M$.
\end{lem}
\proof 
Suppose that $M=0$.
In the case of $M>0$, the proof is parallel with that of $M=0$, but rather easier.

{\bf (Step 0)} 
Let $\Psi\in \hhhh$. 
Let  $H_0=|p-A|+\hf $.
We have 
\begin{align*}
\|  H\Psi\| ^2&=\|  H_0\Psi\| ^2+\|  V\Psi\| ^2 + 2\Re(H_0\Psi,V\Psi),\\
\|  H_0 \Psi\| ^2 &= 
\|  |p-A|\Psi\| ^2 + \|  \hf \Psi\| ^2 + 2\Re (|p-A|\Psi,\hf \Psi).
\end{align*}
Then 
\eq{hiro2014}
\|  H\Psi\| ^2=
\|  |p-A|\Psi\| ^2 + 
\|  \hf \Psi\| ^2 + 
2\Re (|p-A|\Psi,\hf \Psi)+
\|  V\Psi\| ^2 + 
2\Re(H_0\Psi,V\Psi). 
\en
We estimate the three terms 
$\|  |p-A|\Psi\| ^2$, 
$\Re (|p-A|\Psi,\hf \Psi)$ 
and $
\Re(H_0\Psi,V\Psi)
$
on the right-hand side of 
\kak{hiro2014} from below.

{\bf (Step 1)} We estimate $\Re (|p-A|\Psi,\hf \Psi)$. 
Since the  operator $|p-A|$ is singular, 
we introduce an artificial positive mass $m>0$ and 
\eq{defof tmm}
\tmm =\sqrt{(p-A)^2+m^2}.
\en
 We fix $m$ throughout. 
Note that $|p-A|-\tmm $ is bounded. Thus 
\eq{s}
|p-A| =\tmm +(|p-A|-\tmm )
\en
 can be regarded as 
a perturbation of $\tmm $, and the perturbation $|p-A|-\tmm $ is bounded.  
We have 
$\Re (|p-A|\Psi,\hf \Psi)
=\Re(\tmm \Psi,\hf \Psi)+\Re((|p-A|-\tmm )\Psi,\hf \Psi) $.
Since $\Psi\in \hhhh$, $\hf\Psi\in D(p^2)\cap D(\hf)$. In particular 
$\hf\Psi\in D(\tmm )$ and then $\hf \Psi\in D(\tmm ^\han)$. Furthermore we show that 
\eq{app}
\tmm ^\han  \Psi\in D(\hf )
\en in Appendix \ref{papp}.
So we can 
see that 
\begin{align*}
&\Re (|p-A|\Psi,\hf \Psi)\\
&=(\tmm ^\han\Psi,\hf \tmm ^\han\Psi) 
 + \Re (\tmm ^\han\Psi,[\tmm ^\han,\hf ]\Psi)
 + \Re((|p-A|-\tmm )\Psi,\hf \Psi) \\
&
\geq \Re (\tmm ^\han\Psi,[\tmm ^\han,\hf ]\Psi)+
 \Re((|p-A|-\tmm )\Psi,\hf \Psi). 
\end{align*}
We estimate $ ((|p-A|-\tmm )\Psi,\hf \Psi)
$. 
Since
$
 \|  (|p-A|-\tmm )\Psi \|  \leq m \| \Psi\|$,
we see that for each  $\epsilon>0$ there exists $C_1>0$ such that
\begin{eqnarray}\label{0.9}
\Re((|p-A|-\tmm )\Psi,\hf \Psi)
\geq -\eps\|\hf \Psi\|^2-C_1\|\Psi\|^2.
\end{eqnarray}
On the other hand 
we estimate 
${\Re(\tmm ^\han \Psi,[\tmm ^\han,\hf ]\Psi)} $. 
Let $\eps>0$ be given. 
Then there exists $C_2>0$ such that 
\begin{align}\label{0.10}
{\Re(\tmm ^\han \Psi,[\tmm ^\han,\hf ]\Psi)} 
&\geq
-c\|  \tmm ^\han\Psi\| \, \| (\hf +\one)\Psi\|  \nonumber\\
&\geq-\eps \|  |p-A|\Psi\| ^2 -\eps\|  \hf \Psi\| ^2-C_2\| \Psi\| ^2.
\end{align}
The first inequality of \kak{0.10} is derived from 
\eq{f1}
\|[\tmm ^\han, \hf]\Psi\|\leq c\|(\hf+\one)^\han\Psi\|
\en with some constant $c>0$. This is 
shown in Appendix \ref{pf1}.
Hence 
we have 
\eq{step 1}
\Re(|p-A|\Psi, \hf \Psi)\geq 
-\eps\||p-A|\Psi\|^2-2\eps\|\hf\Psi\|^2-(C_1+C_2)\|\Psi\|.
\en

{\bf (Step 2)} We estimate 
$\Re(H_0\Psi,V\Psi)$. 
For each $\eps>0$ there  exists $C_3>0$ such that 
\begin{align*}
\Re(H_0\Psi,V\Psi)&=
\Re((H_0-\tmm -\hf)\Psi,V\Psi)+\Re(\tmm \Psi, V\Psi)+(\hf\Psi, V\Psi)\\
&\geq -\eps\|V\Psi\|^2-C_3\|\Psi\|^2+
\Re(\tmm \Psi, V\Psi).
\end{align*}
We also see that 
$$
 \Re(\tmm \Psi,V\Psi)
 =
 (\tmm ^{1/2}\Psi,V\tmm ^{1/2}\Psi)+
 \Re( \tmm ^{1/2}\Psi,[\tmm ^{1/2},V]\Psi) 
 \geq 
 \Re( [\tmm ^{1/2},V]\Psi,V \Psi).
$$
Recall the 
integral representation
$$
\d \tmm ^{1/2}=\frac{1}{\sqrt{2}\pi}\int_0^\infty \frac{1}{w^{3/4}} 
\left( \tmm ^2+w\right)^{-1}\tmm ^2 {\rm d}w,$$
commutation relations
$$
[(\tmm ^2+w)^{-1}\tmm ^2, V]=
(\tmm ^2+w)^{-1}[\tmm ^2,V]-(\tmm ^2+w)^{-1}
[\tmm ^2,V](\tmm ^2+w)^{-1}\tmm ^2, 
$$
and facts
$$
\d [\tmm ^2,V]
=-2i \sum_{\mu=1}^{d}(p_\mu - A_\mu) 
 V_\mu 
 +\sum_{\mu=1}^{d}
V_{\mu\mu},$$
where $V_\mu=\partial_\mu V$ and $V_{\mu\mu}=\partial_\mu^2 V$. 
Then we have 
\begin{align}
|([\tmm ,V]\Psi, \Phi)|
&=
\left|
\frac{1}{\sqrt{2}\pi}\int_0^\infty \frac{1}{w^{3/4}} 
[\left( \tmm ^2+w\right)^{-1}\tmm ^2, V]\Psi,\Phi){\rm d}w\right|\non \\
&
\label{gaman}
\leq \frac{\sqrt{2}}{\pi} \| \Psi\| \| \Phi\|  \int_0^\infty \frac{{\rm d}w}{w^{3/4}}\sum_{\mu=1}^d
\left(
\frac{
2\|V_\mu\|_\infty
 }{\sqrt{w+m^2}}+
 \frac{\| V_{\mu\mu}\|_\infty}{w+m^2}
 \right).
\end{align}
Thus for each $\eps>0$ there exists $C_4>0$ such that 
\eq{step 2}
\Re(H_0\Psi,V\Psi)\geq   
-\eps\|V\Psi\|^2-C_4\| \Psi\| ^2.
\en

{\bf (Step 3)}
We estimate 
$\||p-A|\Psi\|$. 
Note that 
$$
\|  p_\mu \Psi\| ^2
= \|  (p_\mu-A_\mu)\Psi \| ^2 + 2\Re(A_\mu\Psi,(p_\mu-A_\mu)\Psi) + \|  A_\mu\Psi\| ^2.$$
For each $\epsilon>0$, 
there exist $C_5>0$ and $C_6>0$ 
such that   
\begin{align*}
|\Re(A\Psi,(p-A)\Psi)|
&\leq \epsilon (\|  |p-A|\Psi\| ^2 + \|  \hf \Psi \| ^2)+C_5\| \Psi\| ^2\\
\|  |p| \Psi\| ^2
&\leq (1+\epsilon) \|  |p-A|\Psi\| ^2 + \epsilon \|  \hf \Psi \| ^2 + C_6 \|  \Psi\| ^2. 
\end{align*}
Hence we have 
\eq{step 3}
\||p-A|\Psi\|^2\geq 
\frac{1}{1+\eps}\||p|\Psi\|^2-\frac{\eps}{1+\eps}\|\hf \Psi\|^2-\frac{C_6}{1+\eps}\|\Psi\|^2.
\en

{\bf (Step 4)}
By \kak{step 1}, \kak{step 2}, \kak{step 3} and 
\kak{hiro2014},  we can see \kak{hiro2015} for $\Psi\in \hhhh$.
Let $\Psi\in D(H)$. Since $H$ is essentially self-adjoint on $\hhhh$, by a limiting argument we can see \kak{hiro2015} for $\Psi\in D(H)$.  
\qed

\noindent {\it Proof of Theorem \ref{main}}:\\
We emphasis the dependence on $M$ by writing $H_M$ for $H$.
Let $M>0$. 
Suppose that $V\in \conf$. 
By Lemma~\ref{relative bound}, $H_M$ is closed on 
$\ms D$. 
Then it implies that $H_M$ is self-adjoint on 
$\ms D$
since it is essentially self-adjoint on 
$\ms D$.
Next suppose that $V\in \rel$. 
Then $\T \ \dot + \ \hf$ is self-adjoint on $\ms D$. Since $V$ is also relatively bounded with respect to 
 $\T \ \dot + \ \hf$ with a relative bound strictly smaller than one. Thus $H$ is self-adjoint on $\ms D$.

Let $M=0$. 
By 
$H_0
=H_M+(H_0-H_M)$ and $
\| (H_0-H_M)\Psi\| \leq M\| \Psi\|$,
$H_0$ is self-adjoint on $\ms D$ 
and essentially self-adjoint on $\HHH_{\rm fin}$ by the 
Kato-Rellich theorem. 
\qed

\section{Translation invariant case}
The momentum operator in $\fff$ is defined by the second quantization of the multiplication by $k_\mu$. 
I.e., $\d {\pf}_\mu=\sum_{r=1}^{d-1} \int k_\mu a^{\dagger r}(k) a^r(k) dk$, $\mu=1,..,d$. 
Let $\tot_\mu =p_\mu\otimes\one+\one\otimes{\pf}_\mu$, $\mu=1,...,d$,  be the total momentum operator, 
and we set 
$\tot=(\tot_1,\cdots,\tot_d)$.  
Let $V=0$. 
 Then we can see that 
$[H, \tot_\mu]=0$ and hence $H$ can be decomposed with respect to the spectrum of $\tot_\mu$. 
Thus 
$H\cong\int_{\BR}^\oplus H_P {\rm d}P$, where $H_P$ is called the fiber Hamiltonian with the total momentum $P\in\BR$.

We can see the explicit form of the fiber Hamiltonian. 
Let 
\eq{m6}
L(P)={(P-\pf-A(0))^2+M^2}.
\en
\begin{prop}\cite[Theorem 2.3 (2), Lemma 3.11]{hir07}
Let $P\in \BR$. Then $L(P)$ is  essentially self-adjoint on $\ms C_0=D(\pf^2)\cap D(\hf)$.  
\end{prop}
Set \eq{lbar}
\bar L(P)=\ov{L(P)\lceil_{\ms C_0}}.
\en
\begin{df}
Let $P\in\BR$. 
We define $H(P)$ by 
\eq{HP}
H(P)=\sqrt{\bar L(P)}\ \dot +\ \hf .
\en
\end{df}

\begin{lem}
It follows that 
\eq{m7}
\T +\hf\cong \int _{\BR}^\oplus 
H(P) {\rm d}P.
\en
\end{lem}
\proof 
We define the unitary operator $U$ 
on $\hhh$ by $(UF)(\cdot)\in \HHH$ for $F(\cdot)\in\HHH$ by 
\eq{unitary}
(UF)(P)=(2\pi)^{-d/2}\int _{\BR} e^{iP\cdot x}e^{-i\pf \cdot x} F(x) dx.
\en
It is shown that 
\eq{yui}
U^{-1}\lk 
\int^\oplus_{\BR} 
\bar L(P) {\rm d}P\rk U= (p-A)^2
\en
in \cite[Theorem 2.3]{hir07}.  
Actually it is shown that 
\eq{m1}
(F, \T^2 G)=\int _{\BR} {\rm d}P\lk 
(UF)(P), \bar L(P) (UG)(P)\rk _\fff
\en
for $F, G\in \hhhh$. 
From \kak{yui}  we see that 
$U^{-1} \lk
\int^\oplus_{\BR} e^{-t\bar L(P)} {\rm d}P\rk U=
 e^{-tT^2}$ for all $t\geq 0$ by \cite[Theorem XIII 85 (c)]{rs4}.
Let $F\in \hhhh$. 
By the formula 
\eq{formula}
\T^{\alpha}=
C_\alpha 
\int_0^\infty (\one-e^{-\lambda \T^2})\frac{{\rm d}\lambda}{\lambda^{1+\alpha/2}},
\en
we can see that 
\begin{align*}
(F, \T F)=
C_1\int_0^\infty \frac{{\rm d}\la}{\la^{3/2}}\int_{\BR}
 {\rm d}P((UF)(P), (\one-e^{-\la \bar L(P)})(UF)(P)). 
\end{align*}
By  Fubini's theorem we have 
\begin{align}
\label{m2}
(F, \T F)=
C_1\int_{\BR}
 {\rm d}P \int\frac{{\rm d}\la}{\la^{3/2}}((UF)(P), 
 (\one-e^{-\la \bar L(P)})(UF)(P)). 
\end{align}
Note that $(UF)(P)\in\ffff$ for each $P\in\BR$. 
Hence $(UF)(P)\in D(\bar L(P))\subset D(\sqrt{\bar L(P)})$, which implies that 
\begin{align}
\label{m3}
(F, \T F)=
\int_{\BR}
 {\rm d}P \lk (UF)(P), 
 \sqrt{\bar L(P)}(UF)(P)\rk. 
\end{align}
By the polarization identity and \kak{m3}
we have 
$$(F, \T G)=
\int_{\BR}
 {\rm d}P 
 \lk(UF)(P), 
 \sqrt{\bar L(P)}(UG)(P)\rk.$$
  Furthermore we see that 
 $$\lk F, \lk \T +\hf\rk  G\rk =
\int_{\BR}
 {\rm d}P \lk(UF)(P), 
 H(P)(UG)(P)\rk,$$
 which implies that 
 \eq{m4}
 \T\ \dot + \ \hf =U^{-1}\lk 
 \int_{\BR}^\oplus 
 H(P) {\rm d}P\rk U
 \en 
 on $\hhhh$. 
Since  $\hhhh$ is a core of the left hand side of \kak{m4}, 
 \eq{m5}
 \T\ \dot + \ \hf \cong 
  \int_{\BR}^\oplus 
H(P) {\rm d}P
 \en
holds true as self-adjoint operators. 
Note that 
$ \T\ \dot + \ \hf=\T+\hf $ on 
$D(|p|)\cap D(\hf)$ and $\T+\hf$ is self-adjoint on $D(|p|)\cap D(\hf)$. Then the lemma follows. 
\qed

Let 
$
\ms C=D(|\pf|)\cap D(\hf)$.
Note that $D(|P-\pf|)=D(|\pf|)$ for all $P\in \BR$. 
The essential self-adjointness of $H(P)$ is established in \cite{hir14}.
\begin{prop}\cite[Corollary 7.2]{hir14}
\label{mikan}
Let $M> 0$. 
Then $H(P)$ is 
essentially self-adjoint on $\ms C$.
\end{prop}

The main result in this section is as follows. 
\begin{theorem}\label{main2}
Let $M\geq 0$. 
Then $H(P)$ is self-adjoint on $\ms C$  
and essentially self-adjoint on $\ffff$. 
\end{theorem}
\proof
The proof is parallel with 
that of $H$. 
We show the outline of the proof.
It can be seen that 
there exists a constant $C>0$ such that 
for arbitrary $\Psi\in\ffff$,  
$$\|\sqrt{(P-\pf-A(0))^2+M^2}\Psi\| \leq C(\|  |P-\pf|\Psi\|+ \|\hf^\han\Psi\|+\|\Psi\|).  $$
Then we can derive that 
\eq{a1}
\|H(P)\Psi\|\leq 
C(\| |P-\pf|   \Psi\|+\|\hf \Psi\|+\|\Psi\|)
\en
for $\Psi\in\ffff$. 
In a similar manner to Lemma \ref{lemma1} from \kak{a1} we can see that  
$\ms C_{1}=\ms C\cap \ffff$ is  a core of $H(P)$  for $M>0$. 
Since $|P-\pf|^2$ and $\hf$ are strongly commutative and positive, it is trivial to see 
that 
\eq{yappa}
\| (|P-\pf|^2+\hf)\Psi\|^2 \geq \|    |P-\pf|^2\Psi  \|^2+\|\hf\Psi\|^2.
\en
Since $\ffff$ is a core of 
$|P-\pf|^2+\hf$,  in a similar manner to Lemma \ref{lemma2} we can see that 
$\ffff$ is also a core of $H(P)$ by \kak{yappa}. 
The key inequality to show the self-adjointness of $H(P)$ is 
\eq{key}
\||P-\pf|\Psi\|^2+\|\hf\Psi\|^2\leq C\|(H(P)+\one)\Psi\|^2
\en 
with some $C>0$ for $\Psi\in \ffff$. 
This is shown by using the inequality 
\eq{key2}
\|[\tmm (P)^\han, \hf]\Psi\| \leq c\|(\hf +\one)^\han\Psi\|, 
\en where 
$\tmm (P)=\sqrt{(P-\pf-A(0))^2+m^2}$.
\kak{key2}  is proven  in Appendix \ref{pkey2}.
Thus by \kak{key} in a similar manner to the proof of Theorem \ref{main} 
we can see that $H(P)\lceil_{\ms C}$ is closed. 
Then $H(P)$ is self-adjoint on $\ms C$ for $M\geq0$. 
\qed

\appendix
\section{Stochastic preliminary}
In this appendix we review  functional integral representations of the semigroup generated by 
semi-relativistic  Pauli-Fierz model. This is established in \cite[Theorem 3.13]{hir14}.  
These representations play an important roles to estimate some commutation relations in 
this  paper. 

\subsection{Semi-relativistic Pauli-Fierz model}
Let $(B_t)_{t\geq 0}$ be the $d$-dimensional Brownian motion defined on a Wiener space with Wiener measure $P^x$ starting from $x$. 
Let  $(T_t)$ be the subordinator on a probability space with a probability measure $\nu$ 
such that 
$\Ebb_\nu [e^{-uT_t}]=e^{-t(\sqrt{2u+M^2}-M)}$.
We denote the expectation with respect to the measure 
$P^x\otimes \nu$ by 
$\Ebb_{P\times \nu}^x[\cdots]$. 
Let $a=(a_1(x),\cdots,a_d(x))$ be electromagnetic fields. 
Then 
the semi-relativistic Schr\"odinger operator 
is defined by 
$h=\sqrt{(p-a)^2+M^2}-M +V$.
Then the Feynman-Kac formula \cite[Chapter 3]{lhb11} yields 
the path integral representation of 
$e^{-t h}$ by 
\eq{fy}
(f, e^{-t h}g)=
\ix \left[e^{-\int_0^tV(B_{T_s}) {\rm d}s} 
e^{-i\int_0^{T_t}a(B_s) {\rm d}B_s}
\ov{f(B_{T_0})}g(B_{T_t})\right].
\en
On the other hand the  semi-relativistic Pauli-Fierz  model is defined by the minimal coupling of $h+\hf$ 
with a quantized radiation field $A$:
$$H=\T \, \dot +\, V\, \dot +\, \hf.$$
We can give the functional integral representation of 
$e^{-tH}$ in \cite[Theorem 3.13]{hir14}.
Let
$${\rm q}(F, G)=\half \sum_{\mu,\nu=1}^d (\hat F_\mu, 
\delta_{\mu\nu}^\perp\hat G_\nu)$$ be the quadratic form 
on $\oplus^d \LR$,
where
$\delta_{\mu\nu}^\perp(k)=\delta_{\mu\nu}-k_\mu 
k_\nu/|k|^2$ denotes the transversal delta function.  
Let $\A(F)$ be a 
Gaussian random variables on a probability space $(Q,\Sigma,\mu)$, which  is indexed by 
$F=(F_1,\cdots,F_d)\in \oplus^d \LR$. 
The  mean of $\A(F)$ is 
zero and the covariance is given by 
$\Ebb[\A(F)\A(G)]= {\rm q}(F, G)$. 
Furthermore 
we introduce the Euclidean version of $\A$.
Let
\eq{t1}
{\rm q}_E(F, G)= 
\half \sum_{\mu,\nu=1}^d 
(\hat F_\mu, 
\delta_{\mu\nu}^\perp\hat G_\nu)
\en
be the quadratic form on 
$\oplus^d L^2(\RR^{d+1})$.
On the right-hand side of \kak{t1}, we  note  that 
$(\hat F_\mu, 
\delta_{\mu\nu}^\perp\hat G_\nu)=
\int _{\RR\times \RR^d}
\ov{\hat F_\mu}(k_0,k)
\delta_{\mu\nu}^\perp(k) 
\hat G_\nu(k_0, k) {\rm d}k_ 0{\rm d}k
$ and $\delta^\perp_{\mu\nu}(k)$ is independent of $k_0$.  
Let $\AA(F)$ be a Gaussian random variables on a probability space $(Q_E,\Sigma_E,\mu_E)$, which  is indexed by 
$F\in \oplus ^d L^2(\RR^{d+1})$.
The  mean of $\AA(F)$ is 
zero and the covariance is given by 
$
\Ebb[\AA(F)\AA(G)]=
{\rm q}_E(F,G)$. 
Let us identify $\hhh$ with $L^2(\BR;\fff)$. Thus 
$\Phi\in\hhh$ can be an $\fff$-valued $L^2$-function on $\BR$, 
$\BR\ni x\mapsto \Phi(x)\in\fff$. 
It is well known that 
there exists the family of isometries 
$J_t: L^2(Q)\to L^2(Q_E)$ ($t\in\RR$) 
and $\j_t:L^2(\BR)\to L^2(\RR^{d+1})$
($t\in\RR$) such that 
$J_t^\ast J_s=e^{-|t-s|\hf}$ and 
$\j_t^\ast \j_s=e^{-|t-s|\omega(-i\nabla)}$. 
\begin{prop}
Let $F, G\in \hhh$. 
Then 
\eq{fy2}
(F, e^{-tH}G)=
e^{-tM}\ix \left[e^{-\int_0^tV(B_{T_s}) {\rm d}s} 
(J_0F (B_{T_0}), 
e^{-i\AA(I[0,t])}J_tG (B_{T_t}))_{L^2(Q_E)}\right].
\en
Here 
$
I[0,t]=\oplus_{i=1}^{d}
\int _0^{T_t} \j_{T^\ast s} \tilde \varphi (\cdot-B_s) {\rm d}B_s^i$
is defined by  the limit of 
$\oplus^d L^2(\RR^{d+1})$-valued stochastic integrals  
of  $\tilde \varphi =(\vp/\sqrt\omega\check{)}$, and 
$T_s^\ast=\inf\{t;T_t=s\}$. 
\end{prop}
\proof 
See \cite[Theorem 3.13 and Remark 3.8]{hir14}.  
\qed

Furthermore
let 
\eq{K}
K={ \half (p-A)^2}
\en
be the kinetic term of the  
Pauli-Fierz model $K+V+\hf$. 
The Feynman-Kac formula of $e^{-tK}$ 
is also established as follows. 
\begin{prop}
\label{y2}
Let $F, G\in \hhh$. 
Then 
it follows that 
\eq{hh1}
(F, e^{-tK}G)=
\iix \left[e^{-\int_0^tV(B_{T_s}) {\rm d}s} 
(F(B_0), e^{-i\A(K[0,t])}
G(B_t))_{L^2(Q)}
\right], 
\en
where 
$
K[0,t]=\oplus_{i=1}^ d \int _0^t  
\tilde \varphi (\cdot-B_s) {\rm d}B_s^i
$
is a  $\oplus^d L^2(\BR)$-valued stochastic integral.
\end{prop}
\proof 
See \cite[(4.20), Theorem 4.8]{hir00} and \cite[(7.3.18)]{lhb11}. 
\qed

\subsection{Semi-relativistic Pauli-Fierz model with a fixed total momentum}
Let 
$H(P)=\sqrt{(P-\pf-A(0))^2+M^2}\ \dot +\ \hf $ be the semi-relativistic Pauli-Fierz model with total momentum $P\in \BR$. The rigorous definition of $H(P)$ is given by \kak{HP}. 
The Feynman-Kac formula of $e^{-tH(P)}$ is also established. 
\begin{prop}
Let $F, G\in L^2(Q)$. Then 
\eq{hiro13}
(F, e^{-tH(P)}G)=
e^{-tM}
\Ebb_P^0 \left[ 
(J_0F (B_{T_0}), 
e^{-i\AA(I[0,t])}e^{i(P-\pf)\cdot B_{T_t}}J_t 
G (B_{T_t}))_{L^2(Q_E)}\right].
\en
\end{prop}
\proof
This is proven by a minor modification of \cite[Theorem 3.3]{hir07}. 
\qed

Furthermore the kinetic term of the Pauli-Fierz model with total momentum $P\in\BR$ 
is given by 
$$K(P)=\half {(P-\pf-A(0))^2},\quad P\in\BR.$$
 The Feynman-Kac formula of $e^{-tK(P)}$ is also established as follows. 
\begin{prop}
Let $F,G\in L^2(Q)$. Then 
\eq{hiro131}
(F, e^{-tK(P)}G)=
\Ebb_P^0 \left[ 
(F(B_{0}), 
e^{-i\A(K[0,t])}e^{i(P-\pf)\cdot B_t}
G (B_{t}))_{L^2(Q)}\right].
\en
\end{prop}
\proof 
This is also proven by a minor modification of \cite[Theorem 3.3]{hir07}. 
\qed

\section{Proof of Lemma \ref{hiroshima}}
\label{phiroshima}
{\it Proof of Lemma \ref{hiroshima}}: \\
It is shown that $e^{-tK}$ leaves $D(p^2)\cap C^\infty(N)$ invariant in \cite[Lemma 7.53]{lhb11}. 
See also \cite[Theorem 2.6]{hir00}. 
It is enough to show that $e^{-tK}D(\hf )\subset D(\hf)$. By the Feynman-Kac formula we have 
\begin{align*}
&(\hf F, e^{-tK} G)=\iix[(\hf F(B_0), e^{-i\A(K[0,t])}G(B_t))]\\
&=
(F, e^{-tK} \hf G)+\iix[(F(B_0), [\hf, e^{-i\A(K[0,t])}]G(B_t))]. 
\end{align*}
We can estimate  as 
$ [\hf, e^{-i\A(K[0,t])}]= e^{-i\A(K[0,t])}(\Pi(K[0,t])+\xi)$, where 
$\Pi(K[0,t])=[\hf, \A(K[0,t])]$ and
$\xi=q(K[0,t],K[0,t])$. 
Thus  we see that
\eq{b1}
\left|
\iix[(F(B_{0}), [\hf, e^{-i\A(K[0,t]) }] G(B_{t}))
]\right|
\leq 
C
(t+\sqrt t)
\|F\| \|(\hf+\one)^\han G\|. 
\en
Here we used  that 
$\|\Pi(K[0,t])\Psi\|\leq C(\|K[0,t]\|+\|K[0,t]/\sqrt\omega\|) \|(\hf+\one)^\han \Psi\|$ and 
BDG-type inequality (\cite[Theorem 4.6]{hir00} and \cite[Lemma 7.21]{lhb11}):
\begin{align}
&
\label{bdg1}
\Ebb_P^0[\xi^2]\leq t^2 C,\\
&
\label{bdg2}
\Ebb_P^0[(\|K[0,t]\|+\|K[0,t]/\sqrt\omega\|)^2]\leq C t.
\end{align}
Then we have 
$$|(\hf F, e^{-tK}G)|\leq 
C (t+\sqrt t)
\|F\|
\|(\hf+\one)^\han G\|+\|F\|\|\hf G\|,$$
and the desired results follow. 
\qed

\section{Proof of Proposition \ref{hiro11}}
\label{phiro11}
\begin{lem}
\label{k1}
Let $V\in \conf$.
Then 
$e^{-tH}$ leaves $D(V)$ invariant, i.e., 
$e^{-tH} D(V)\subset D(V)$.
\end{lem}
\proof
Let $F,G\in D(V)$. 
We define $Q_{[0,t]}$ by 
$Q_{[0,t]}=e^{-tM}e^{-\int_0^tV(B_s)ds}J_0^\ast e^{-i\AA(I[0,t])}J_t:\hhh\to\hhh$. Then we have 
$$(VF, e^{-tH}G)=\ix[(V(B_{T_0})F(B_{T_0}), Q_{[0,t]}G(B_{T_t}))].$$
Hence
we see that 
\begin{align*}
(VF, e^{-tH}G)=
(F, e^{-tH} V G)
+
\ix[(F(B_{T_0}), Q_{[0,t]}(V(B_0)-V(B_{T_t}))G(B_{T_t}))]
\end{align*}
and, by the Taylor expansion $V(x)-V(B_{T_t}+x)=\sum_\mu 
(\partial_\mu V(\xi)) B_{T_t}^\mu$ with some $\xi\in\BR$, 
we can estimate as 
$$
\left|
\ix[(F(B_{T_0}), Q_{[0,t]}
(V(B_0)-V(B_{T_t}))
G(B_{T_t}))]
\right|
\leq \|F\|\||x|G\|\sup_x\sqrt{
\sum_\mu|\partial_\mu 
V(x)|^2}.$$
Here we used the fact that $G\in D(|x|)$. 
Then we have 
$$|(VF, e^{-tH}G)|\leq C\|F\| (\||x| G\|+\|VG\|)$$ with some constant $C>0$. Then 
$e^{-tH}G\in D(V)$ follows. 
\qed

{\it Proof of Proposition \ref{hiro11}}: \\
Suppose that $V$ satisfies (2) of Assumption \ref{h2}.  
It is shown in \cite[Lemmas 4.3 and 4.4]{hir14}
that 
$D(H)\subset \cap_\mu D(p_\mu) \bigcap D(\hf)$ and 
$e^{-tH}$ leaves $\cap_\mu D(p_\mu)\bigcap D(\hf)$ invariant, which 
implies that  
$e^{-tH}$ leaves $D(|p|)\cap D(\hf)$  invariant. 
Combining this with Lemma \ref{k1} we see that 
$D(H)\subset \ms D$ 
and 
$e^{-tH}$ leaves $\ms D$ invariant.
Then $\ms D$ is a core of $H$ by \cite[Theorem X.49]{rs2}.
\qed

\section{Proof of \kak{app}}
\label{papp}
Note that $\tmm ^\han=(2K+m^2)^{1/4}$, 
where $K$ is given by \kak{K}. 
We have 
\eq{y1}
(2K+m^2)^{\alpha/2}=
C_\alpha 
\int_0^\infty (\one-e^{-\lambda (2K+m^2)})\frac{{\rm d}\lambda}{\lambda^{1+\alpha/2}}
\en
for $0\leq\alpha< 2$ with some constant $C_\alpha$.  
From this formula we have the lemma below:
\begin{lem}
\label{k2}
There exists $C>0$ such that 
$$(F, \tmm ^\han G)=
C\int_0^\infty 
\lkk 
(F, G)-e^{-\lambda m^2/2} \iix[(F(B_0), 
e^{-i\A(K[0,\la]) } G(B_\la))
]\rkk \frac{{\rm d}\lambda}{\lambda^{5/4}}.$$
\end{lem}
\proof 
This can be derived from 
Proposition \ref{y2},  \kak{y1} and changing the variable.  
\qed

\noindent
{\it Proof of \kak{app}}: \\
Let $F\in D(\hf)$ and $G\in \hhhh$. 
Thus $\hf G\in \hhhh$. 
By \kak{y1} we have 
$$
(\hf F, \tmm ^\han G)=
C\int_0^\infty \!\!\! 
\lkk 
(\hf F, G)-e^{-\lambda m^2/2}
\!\!\! \iix[(\hf F(B_{0}), e^{-i\A(K[0,\la]) } G(B_{\la}))
]\rkk \frac{{\rm d}\lambda}{\lambda^{5/4}}.$$
Then 
we have 
\begin{align}
&(\hf F, \tmm ^\han G)-
(F, \tmm ^\han \hf G)\non\\
&\label{y3}
=-C
\int_0^\infty 
\frac{e^{-\lambda m^2/2} }{\lambda^{5/4}} {\rm d}\lambda\iix[(F(B_{0}), [\hf, e^{-i\A(K[0,\la]) }] G(B_{\la}))
].
\end{align}
We have 
$$[\hf, e^{-i\A(K[0,\la]) }]=e^{-i\A(K[0,\la]) }(\Pi(K[0,\la])+\xi),$$
where 
$\Pi(K[0,\la])=[\hf, \A(K[0,\la])]$ and
$\xi=q(K[0,\la],K[0,\la])$. 
Thus  we see that in a similar manner to \kak{b1}, \kak{bdg1} and \kak{bdg2},  
\begin{align}
&\left|
\int_0^\infty 
\lkk 
\iix[(F(B_{0}), [\hf, e^{-i\A(K[0,\la]) }] G(B_{\la}))
]\rkk \frac{e^{-\la m^2/2} {\rm d}\lambda}{\lambda^{5/4}}\right|\non 
\\
&\label{anata}
\leq 
C\int_0^\infty 
\frac{\sqrt \lambda+\lambda}{\lambda^{5/4}}e^{-\lambda m^2/2} 
{\rm d}\lambda \|F\| \|(\hf+\one)^\han G\|. 
\end{align}
Then we see that 
$|(\hf F, \tmm ^\han G)|\leq 
C\|F\| \|(\hf+\one)  G\|$ with some constant $C>0$. 
Hence $\tmm ^\han G\in D(\hf)$ follows. \qed

\section{Proof of \kak{f1}}
\label{pf1}
 {\it Proof of \kak{f1}}:\\
The proof of \kak{f1} is similar to that of 
\kak{app}. 
Let $G\in \hhhh$. 
By \kak{y3} and \kak{anata}, 
it follows that 
$|(F, [\hf, \tmm ^\han]G)|\leq C \|F\|\|(\one+\hf)^\han G\|$. This implies \kak{f1}. 
\qed

\section{Proof of \kak{key2}}
\label{pkey2}
{\it Proof of \kak{key2}}: \\
The idea of the proof of \kak{key2} is similar to 
\kak{app} and \kak{f1}.
We have 
\begin{align*}
( \Phi, [\hf, \tmm (P)^\han] \Psi)=
( \hf \Phi,  \tmm (P)^\han \Psi)-
( \Phi,  \tmm (P)^\han\hf \Psi)
\end{align*}
The Feynman-Kac formula yields that 
\begin{align*}
( \Phi, [\hf, \tmm (P)^\han] \Psi)=
\!\!\int_0^ \infty \!\!\frac{e^{-m^2\la/2} {\rm d}\la}{\la^{5/4}}
\Ebb_P^0[e^{iP\cdot B_\la}
(\Phi(B_0), 
[\hf, e^{-i\A(K[0,\la])}]e^{-i\pf\cdot  B_\la}\Psi(B_\la))]. 
\end{align*}
Since 
$[\hf, e^{-i\A(K[0,\la])}]=e^{-i\A(K[0,\la])}(\Pi(K[0,\la])+\xi)$. 
Then in a similar manner to 
\kak{anata} we can derive the desired results. 
\qed

\end{document}